\documentclass[10pt]{article}

\usepackage{amsmath}
\usepackage{amssymb}

\usepackage{graphicx}
\usepackage{natbib}

\usepackage[usenames]{color} 

\usepackage[labelfont=bf,labelsep=period,justification=raggedright]{caption}

\def\R{\mathbb{R}}
\def\T{{\cal T}}
\def\J{{\cal J}}

\begin{document}

\begin{flushleft}
{\Large
\textbf{Complexity and anisotropy in host morphology make populations safer against epidemic outbreaks}
}
\\
\vskip10pt
Francisco~J.~P{\'e}rez-Reche$^{1,\ast}$,
Sergei~N.~Taraskin$^{2}$,
Luciano~da~F.~Costa$^{3}$,
Franco~M.~Neri$^{4}$,
Christopher~A.~Gilligan$^{4}$
\\
\bf{1}  Department of Chemistry, University of Cambridge, Cambridge, UK
\\
\bf{2} St. Catharine's College and Department of Chemistry,
University of Cambridge, Cambridge, UK
\\
\bf{3} Institute of Physics at S\~ao Carlos, University of
S\~ao Paulo, P.O. Box 369, S\~ao Carlos, S\~ao Paulo,
13560-970, Brazil, and National Institute of Science and Technology of Complex Systems, Brazil
\\
\bf{4} Department of Plant Sciences, University of Cambridge,
Cambridge,  UK
\\
$\ast$ E-mail: fjp23@cam.ac.uk
\end{flushleft}

\section*{Abstract}
One of the challenges in epidemiology is to account for the complex
morphological structure of hosts such as plant roots, crop fields,
farms, cells, animal habitats and social networks, when the
transmission of infection occurs between contiguous hosts.
Morphological complexity brings an inherent heterogeneity in
populations and affects the dynamics of pathogen spread in such
systems.  We have analysed the influence of realistically complex host
morphology on the threshold for invasion and epidemic outbreak in an
SIR (susceptible-infected-recovered) epidemiological model.  We show
that disorder expressed in the host morphology and anisotropy reduces
the probability of epidemic outbreak and thus makes the system more
resistant to epidemic outbreaks.  We obtain general analytical
estimates for minimally safe bounds for an invasion threshold and then
illustrate their validity by considering an example of host data for
branching hosts (salamander retinal ganglion cells).  Several spatial
arrangements of hosts with different degrees of heterogeneity have
been considered in order to analyse separately the role of shape
complexity and anisotropy in the host population.  The estimates for
invasion threshold are linked to morphological characteristics of the
hosts that can be used for determining the threshold for invasion in
practical applications.

\section{Introduction}
One of the main questions in epidemiology regards the
outbreak of epidemics, i.e. whether an infectious disease can spread
throughout a given ensemble of hosts or not. Many systems display a
threshold for epidemics which divides the parameter space into regions
where an outbreak may occur from regions where the disease cannot
spread \citep{Murray_02:book,Marro_99:book}.  Identification of the
factors that determine such a threshold is of great importance in
devising robust strategies to control the spread of disease through a
population of susceptible hosts. Successful control deflects the
system into the non-invasive region of parameter space so that a
pathogen fails to invade. This simple concept can be applied to the
spread of infection and disease, at a range of scales from the
cellular, in which the host comprises a single cell, through
populations of plants and animals in which a host equates with an
individual organism, up to larger-scale systems, in which the unit of
interest may be an individual field or farm for crop and livestock
disease, a school, a village or other natural clustering for human
disease.  The challenge at each scale lies in dealing in a
quantitative manner with factors such as stochasticity and
heterogeneity, so that thresholds used to identify strategies for
control are robust to these uncertainties. Stochasticity in disease
spread is linked to the fact that susceptible hosts become infected
only with a certain probability when challenged by inoculum from
infected hosts under otherwise identical circumstances.
Heterogeneity, on the other hand, is associated with a range of
factors that may differ amongst hosts. These include disorder in
characteristics such as pathogen infectivity and host susceptibility,
and the spatial arrangement of hosts. For certain types of hosts,
heterogeneity also includes the inherent morphological complexity
(i.e. irregularity in shape) of
the host. Morphological complexity is especially important in
epidemiological spread where it affect the contact rate between
contiguous hosts.  Examples of morphologically complex hosts include
dendritic cells and plants, crop fields and farms, and individual
clusters in social networks
\citep{Davis_Nature2008,Gonzalez_Nature2008,Soriano_PNAS2008,Boender2007_Plos,Eisinger2008}.

While the problem of disease invasion has been extensively studied,
both experimentally and theoretically, most attention has been focused
first on deterministic systems and increasingly on stochastic models
\citep{Murray_02:book,Truscott_PNAS2003,Gibson_PNAS2004}. Heterogeneity
has traditionally received less attention although there are some
remarkable exceptions (e.g. 
\cite{Levin1996,Newman2002,sander2002,sander2003,Cook_PNAS2007,Boender2007_Plos,Eisinger2008,Miller2007,Kenah2007,Volz2008}
).
However, none of the previous work has established a link
  between heterogeneity and morphological features of the system leading to such a heterogeneity.
In particular, the heterogeneity associated with
  the host morphology and the effects it may have on the features of
  the epidemics remain to be understood. 

\begin{figure}[h]
  \begin{center}


{\includegraphics[clip=true,width=8.4cm]{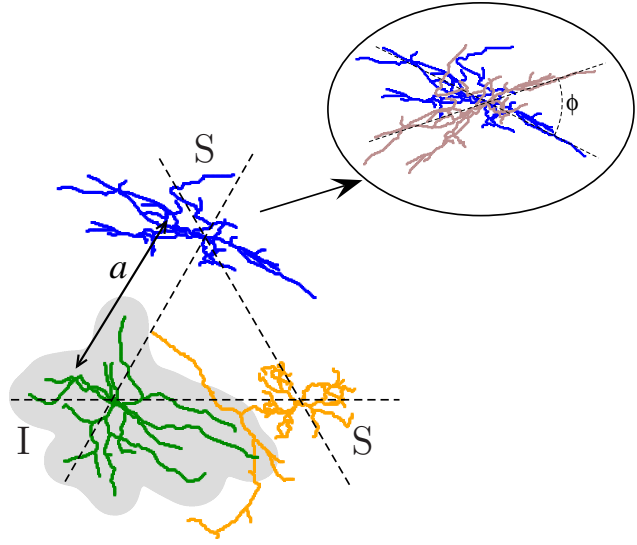}}

\caption{Example of a
  system formed by complex hosts represented,
  for concreteness,
  by planar neurons
  corresponding to the salamander retinal ganglion cells
  placed on the nodes of a triangular lattice with   lattice
  spacing $a$ in such a way
  that the somata coincide with the
  lattice nodes. The pathogen infests the surroundings (shaded area) of
  infected (I) hosts and, eventually, reaches the neighbouring
  susceptible hosts
  (e.g. amber susceptible (S) neuron on the right from I).
  The probability of infection of a susceptible host and the probability of
  a global epidemic outbreak   depend on
  overlaps, $J$, between the infested region and susceptible hosts,
  and infection efficiency, $k$.
  The overlaps are dictated by the host morphology and spatial
  arrangement of hosts.
  The infection efficiency determines the effectiveness of the contact
  in terms of transmission of infection.
  Besides these factors, the inset shows that $J$,
  and thus the invasion
  threshold depend, in general, on the local orientation of the hosts,
  $\phi$.}
\label{fig:neurons}
  \end{center}
\end{figure}

In systems  where the pathogen is transmitted between hosts due to
their proximity, one can identify three main factors that determine
the invasion threshold (see Fig.~\ref{fig:neurons}):
(i) spatial arrangement of the hosts in the population;
(ii) \emph{morphology}, i.e. the shape of the hosts, and
(iii) the infection efficiency resulting from the
net effect of the interplay between the pathogen infectivity and the
host susceptibility upon contact.

Several models have been proposed for the dynamics of epidemics
spreading by contacts between hosts
\citep{Murray_02:book,Liggett_85:book,Marro_99:book,Hinrichsen_00,Odor_04:review}.
In such models, the hosts can be in different states, e.g.
susceptible (S), infected (I), and recovered (or removed) (R) in the
prototype SIR epidemiological model.  The state of each host may
change according to certain model-dependent rules.  For instance, the
SIR model assumes that infected hosts can infect others that are
susceptible and then become recovered and fully immune to further
pathogen attacks.  In such models, the stochasticity and heterogeneity
for the spreading process are treated in a simplistic way by using
phenomenological probability densities for relevant parameters and
thus not linking the host morphology and invasion threshold.

In this paper, we establish a quantitative link between host
morphology and the invasion threshold in an ensemble of hosts with
realistically complex morphology.  By analysing the conditions for
epidemic outbreak in several systems with different degrees of
configurational heterogeneity, we conclude that the invasion threshold
is mainly determined by: (i) the average overlap between neighbouring
hosts, (ii) the morphological complexity of hosts, and (iii) the host
shape anisotropy.  In particular, we demonstrate analytically and
numerically that the resilience of the system to invasion increases
with morphological complexity and anisotropy of hosts.
This result is valid under very general
  conditions and is therefore applicable to a wide range of host
  ensembles.
We show that
irrespective of the degree of the host anisotropy the spreading
process can be described in terms of a mean-field system, so that
analytical estimates for the invasion threshold can be obtained. In
addition, we complete our analysis by identifying several
morphological characteristics that can be used for determining the
threshold for invasion resulting in an epidemic outbreak.  
%

\section{Methods}
\label{sec:methods}
We consider a set of $N$ morphologically different branching structures, $n=1,2,\dots N$, placed on all nodes
$i=1,2,\dots,L^2$ of an $L \times L$ regular lattice with spacing $a$
and nearest neighbour links only
(see Table \ref{Table_Symbols} for a summary of the
  notation used in the text).
In particular,
we deal with a triangular lattice (see Fig.~\ref{fig:neurons}) because this arrangement corresponds to the
organisation of hosts with the highest risk of epidemic spread, in the sense that any other 2D regular lattices with
the same lattice spacing is more resistant to epidemic invasion  \citep{Isichenko_RMP1992,Stauffer1994}. An additional
advantage of the triangular lattice is that the next-nearest-neighbour links (ignored in our approach) are less likely
in such a lattice as compared with other 2D lattices. A certain spatial configuration of generally anisotropic
branching structures is fully defined in terms of the set of hosts placed on the lattice nodes, $\{n_i\}$ (where $n_i$
is the host number placed on node $i$), and their orientations, $\{\phi_i\}$. The system of hosts can be either
homogeneous, if morphologically identical hosts, $n_i = n$, of the same orientation, $\phi_i = \phi$, are placed on the
lattice, or heterogeneous if the hosts $n_i$ and/or their orientations $\phi_i$ are chosen, e.g. at random.
Specifically, we consider three types of arrangements with different degrees of heterogeneity which allow the various
factors contributing to the invasion threshold to be analysed separately:
\begin{itemize}
\item[*]{Arrangements 1:} both $n_i$ and $\phi_i$ are distributed
  according to uniform distributions so that all possible hosts and
 orientations are equally probable.
 These are highly disordered configurations.

\item[*]{Arrangements 2:} the same host is placed on all the nodes
  with the same orientation (i.e., $n_i=n$ and $\phi_i=\phi$ for all
  the nodes $i$).  Such ordered arrangements highlight the role of the
  host anisotropy leading to different overlaps along different lattice
  directions.

\item[*]{Arrangements 3:} the same host is placed on all the nodes and
  its orientation is drawn from a uniform distribution
  of width  $\Delta \phi$ and mean value $\bar{\phi}$.
  Such arrangements allow both the morphological complexity and
  anisotropy of hosts to be analysed in a comparative manner and they
  contain arrangements of type 2 as particular cases with
$\bar{\phi}=\phi$ and $\Delta \phi =0$.
\end{itemize}


\begin{table}[h]
\label{Table_Symbols}
\caption{Symbols and definitions used in the text. Indexes $i$ and $j$
span the $L \times L$ nodes in the (triangular) lattice. Index
$\alpha$ spans the three main lattice directions, i.e.,
$\alpha=1,2,3$. Hosts are labeled by an index $n=1,2,\dots,N$. The
symbol $p_c$ stands for the bond percolation threshold.}
\begin{center}
\begin{tabular}[t]{p{6cm} p{8.5cm}}
\hline
\multicolumn{2}{l}{\bf Host characteristics, arrangement, and overlaps}  \\
\hline
$\rho({\mathbf r};n)$ & Host density \\
$\phi$ & Host orientation \\
$a$ & Lattice spacing\\
$J_{ij}=\int \rho({\mathbf r};n_i)
\rho({\mathbf r}-{\mathbf a};n_j)
\text{d}{\mathbf r}$ & Overlap between hosts at nodes $i$ and $j$\\
$\J=\{J_{ij}\}$ & Set of overlaps between all the pair of hosts\\
$V_1=\frac{1}{3}\sum_{\alpha=1}^3 \left(\langle {\cal{J}}_{\alpha}^2\rangle - \langle
{\cal{J}}_{\alpha}\rangle^2\right)$ & Dispersion of $\J$ associated
with the shape complexity\\
$V_2=\frac{1}{3(1+p_c)}\sum_{\alpha=1}^3 (\langle {\cal{J}}_{\alpha}\rangle -
\langle {\cal{J}} \rangle)^2$ & Dispersion of $\J$ associated
with the shape anisotropy\\
\hline
{\bf Disease transmission} & \\
\hline
$k$ & Infection efficiency\\
$\tau=1$ & Recovery time \\
$\beta_{ij}=k J_{ij}$ & Transmission rate \\
$T_{ij}=1-e^{-\beta_{ij}\tau}=1-e^{-k J_{ij}}$ & Transmissibility \\
\hline
{\bf Invasion of infection} & \\
\hline
$P_{\text{inv}}(k,a)$ & Probability of invasion\\
$a_c(k)=\inf\{a:P_{\text{inv}}(k,a)=0\}$ & Invasion threshold in
terms of the lattice spacing\\
$k_c(a)=\sup \{k:P_{\text{inv}}(k,a)=0\}$ & Invasion threshold in
terms of the infection efficiency \\
$k_c^0 \propto \langle \J \rangle^{-1}$ & Invasion threshold in a mean-field system with overlaps
$\langle \J \rangle$\\
$\Delta k_1 \propto V_1$ & Non-mean-field contribution to $k_c$
originated by the shape complexity\\
$\Delta k_2 \propto V_2$ & Non-mean-field contribution to $k_c$
originated by the shape anisotropy\\
\hline
\multicolumn{2}{l}{\bf Morphological characteristics relevant to the invasion threshold}\\
\hline
$r^{\text{mf}}$ & Radius of effective circular hosts in the mean-field
approximation\\
$\{d_1,d_2,d_3\}$, ($d_1 \le d_2 \le d_3$) & Lattice-adapted diameters\\
SLAD $=d_2$ & Second lattice-adapted diameter\\
\hline
\end{tabular}
\end{center}
\end{table}

In order to investigate the spread of epidemics in such systems, we
apply the dynamical rules of the SIR formalism.  Infection and hence
disease can be transmitted between infected and susceptible hosts with
transmission rate $\beta$, and infected hosts recover after a fixed
time $\tau$.  We assume that the value of $\tau$ is time-independent,
identical for all the hosts, and thus can be chosen as the time scale
of the problem, $\tau=1$.  Such homogeneity in $\tau$ provides a
minimally safe bound for the invasion threshold
\citep{kuulasmaa1982,Cox1988} (see brief explanation in Appendix
\ref{App.Homo_tau}).

The transmission of infection from an infected host, $n_i$ at node
$i$, to a susceptible nearest neighbour, $n_j$ at node $j$, separated
by a unit-cell vector ${\mathbf a}$, is a Poisson process with a
transmission rate $\beta_{ij}$.  The value of $\beta_{ij}$ is assumed
to be proportional to the overlap $J_{ij}$ between hosts $i$ and $j$,
i.e.  $\beta_{ij} = k J_{ij}$, where $k$ is the infection efficiency
which accounts for the effectiveness of the overlap for
transmission of infection. We shall set $k$ to be identical for all the pairs of
nearest neighbours.  Possible variability in $k$ can be easily
incorporated into the model but it does not change the main results
qualitatively.

The overlap between hosts is defined as
\begin{eqnarray}
J_{ij} = \int \rho({\mathbf r};n_i)
\rho({\mathbf r}-{\mathbf a};n_j)
\text{d}{\mathbf r}
~,
\label{eq:overlap_int}
\end{eqnarray}
in terms of the host density,
\begin{equation}
\rho({\mathbf r};n_i)=\sum_{{\mathbf p}\in n_i} \delta({\mathbf
  r}-{\mathbf p}) ~,
\label{eq:density}
\end{equation}
where the position vector ${\mathbf p}$ scans all the points in the host
structure $n_i$.

One of the main quantities involved in the SIR process is the
transmissibility $T_{ij}$ defined as the probability that the pathogen
is transmitted from an infected host at node $i$ to infect a
susceptible host located at node $j$ during the life-time, $\tau$, of
the infected node. For a Poisson process, the transmissibility is
given by the following expression \citep{grassberger1983},
\begin{equation}
T_{ij}=1-e^{-\beta_{ij}\tau}=1-e^{-k J_{ij}}~.
\label{eq:Tij_def}
\end{equation}
Therefore, the value of $T_{ij}$ depends on the lattice spacing, $a$, which
determines the overlap, $J_{ij}$,  and the infection efficiency, $k$.

In a finite-size population, the SIR process lasts for finite
  time and after its termination two types of hosts, R and S, can be
  found in the system. The region of the recovered hosts can be mapped
  onto one of the clusters in the bond percolation problem by mapping
  the transmissibilities $\left\{T_{ij}\right\}$ to the bond
  probabilities \citep{grassberger1983}. According to this mapping, the
  probability of invasion of disease starting from a single host,
  $P_{\text{inv}}(k,a)$, is identified with the probability that this
  first infected host belongs to the infinite cluster of connected
  sites in the bond-percolation problem.  The invasion threshold is
  formally defined as the boundary between the invasive and the
  non-invasive regimes characterised by $P_{\text{inv}}>0$ and
  $P_{\text{inv}}=0$, respectively. This condition introduces a
  separatrix in the parameter space which can be given in terms of
  several (control) parameters of the system, such as, for example, the
  infection efficiency and lattice spacing. In terms of the infection
  efficiency, the invasion threshold is the value $k_c = \sup
  \{k:P_{\text{inv}}=0\}$ such that relatively small values of $k \le
  k_c$ correspond to non-invasive regime (i.e., $P_{\text{inv}}=0$),
  while larger values, $k > k_c$, describe the invasive domain (i.e.,
  $P_{\text{inv}}>0$).  Similarly, the critical lattice spacing,
  $a_c =\inf\{a:P_{\text{inv}}=0\}$, splits the range of lattice
  spacings into two regions: $a<a_c$ and $a \geq a_c$ corresponding to
  invasive and non-invasive regimes, respectively. The host morphology
  affects implicitly (through the overlaps between hosts $\{J_{ij}\}$)
  both $k_c(a)$ and $a_c(k)$. The morphological variability of the
  hosts together with the heterogeneity in their arrangement makes the
  set $\{J_{ij}\}$ disperse in general. The overlap can then be
  regarded as a random variable ${\cal{J}}$ taking the values
  $\{J_{ij}\}$. We will describe ${\cal{J}}$ in terms of its average,
  $\langle {\cal{J}} \rangle$, and deviations from $\langle {\cal{J}}
  \rangle$ originated by its dispersion. Consequently, it is
  convenient to split the expression for invasion threshold into two
  contributions:
\begin{equation}
\label{eq.General_kc}
k_c=k_c^0+\Delta k,
\end{equation}
where $k_c^0$ and $\Delta k$ are associated with the average and the dispersion of ${\cal{J}}$, respectively.
 For the spatial arrangements of branching hosts considered below, the source of dispersion in the
overlaps is two-fold: (i) $V_1$, which is due to variability arising
from complexity in the shapes of different hosts for arrangements of
type 1 and (ii) $V_2$, which is due to variability in shape anisotropy
for arrangements of type 2. Correspondingly, the value of $\Delta k$
can be split into two components, i.e.  $\Delta k=\Delta k_1 +\Delta k_2  $, where $\Delta k_1 \propto V_1$  and
$\Delta k_2 \propto V_2$. Both complexity  and anisotropy contribute to dispersion of the overlaps for arrangements of
type 3.

For numerical illustration and concreteness, in this paper we use a
set of $N$ ($N=51$) neurons (Fig.~\ref{fig:neurons}) corresponding to
the salamander retinal ganglion cells \citep{Ascoli2006}, which are
mostly planar, as typical representatives of complex branching
structures. In this case, the vector ${\mathbf p}$ introduced in
  Eq.~(\ref{eq:density}) scans all the pixels defining the digital
  image of each neuron.  Technically, the $\delta$-functions in
  Eq.~(\ref{eq:density}) are replaced by Gaussians of width comparable
  to the pixel size.  This broadening mimics the (diffusive) spreading
  of the pathogen around the host. While we are not aware of
documented examples of pathogen transmission in these structures, they
are representative of a broad class of structures that are known to
transmit virus infections (e.g.
\cite{LaVail:1997,Ehrengruber_JVirol2002,Chen_JVirol2007,Samuel:2007}
). The importance here is to use the published data
on complex morphology to test the general methods introduced below.

\section{Results}
\label{sec:results}

In this section, we analyse the invasion threshold resulting in each
of the spatial arrangements listed above
and propose morphological characteristics for description of the
invasion threshold.

\subsection{Arrangements 1. $n_i$ and $\phi_i$ random}

\begin{figure}[h]
\begin{center}


{\includegraphics[clip=true,width=8.4cm]{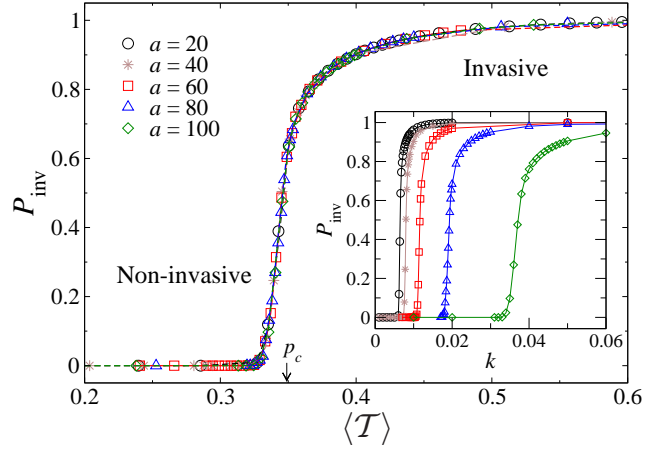}}


  \caption{Probability of invasion, $P_{\text{inv}}$, for
    morphologically complex hosts in disordered arrangements of type 1
    on a lattice of size $L\times L = 200\times 200$.  The main plot
    displays $P_{\text{inv}}$ as a function of the average
    transmissibility, $\langle {\cal{T}} \rangle$, for heterogeneous
    systems with different lattice spacings $a$ (marked by different
    symbols) and for a mean-field system with homogeneous
    transmissibility $\langle {\cal{T}} \rangle$ (dashed line).  The
    bond-percolation critical probability, $p_c$, marked by arrow
    gives the invasion threshold in the thermodynamic limit.  The
    inset shows the invasion probability as a function of the
    infection efficiency $k$ for different values of lattice spacing
    marked by the same symbols as in the main figure.  }
\label{fig:Dis_1}
\end{center}
\end{figure}

\begin{figure}
  \begin{center}


\begin{tabular}{c}
{\includegraphics[clip=true,width=8.4cm]{Fig3_a.eps}}
\\
{\includegraphics[clip=true,width=8.4cm]{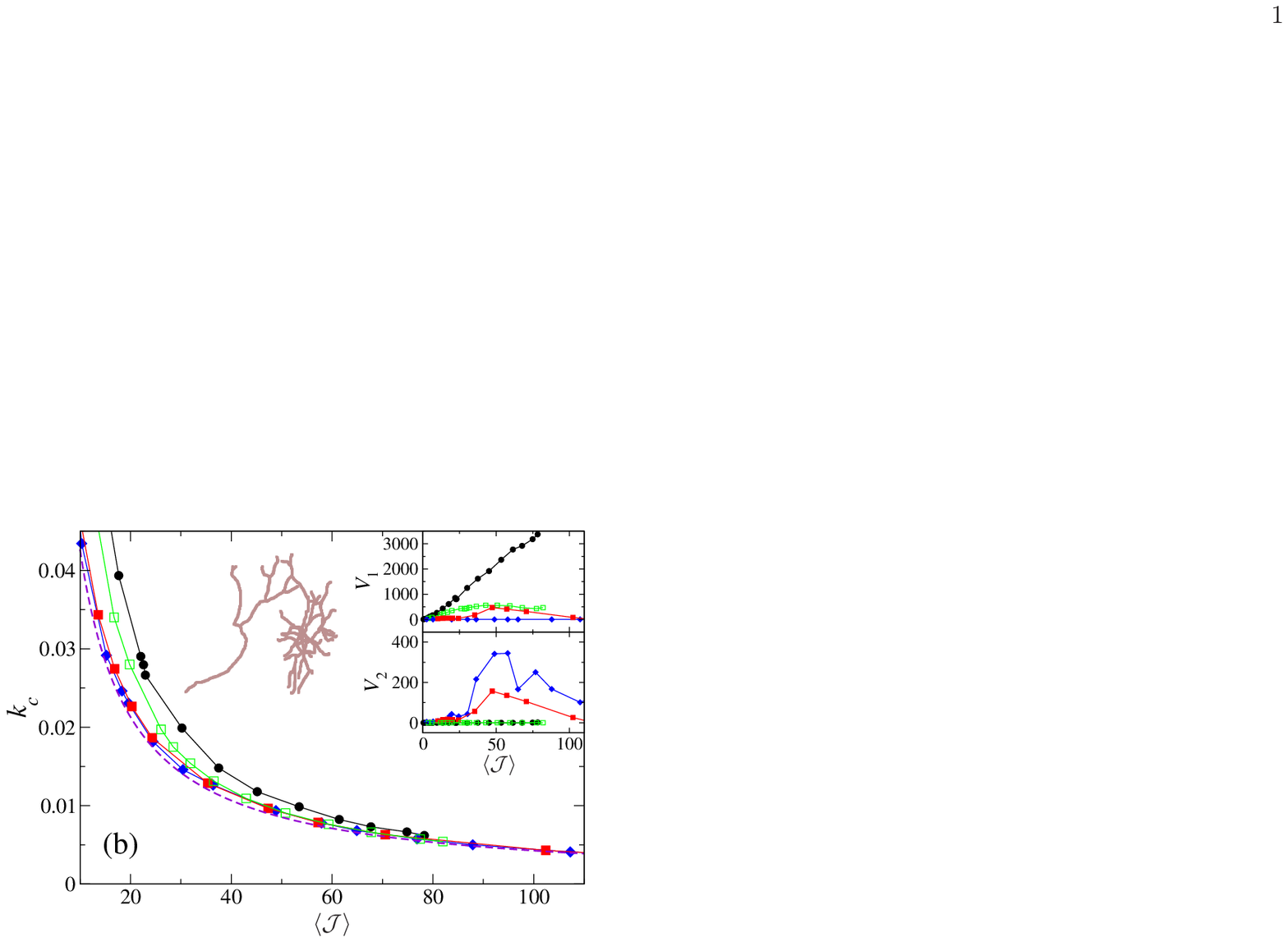}}\\
\end{tabular}


\caption{ Invasion threshold for the system of branching hosts.  (a)
  Representation in terms of the infection efficiency, $k$, and the
  lattice spacing, $a$. The threshold $k_c(a)$ for arrangements of
  type 1 is shown
  by circles. The shaded region   corresponds to   the
  statistically 
possible values for the estimation of $k_c(a)$ in terms of effective
  circular hosts with homogeneous overlaps defined in
  Eq.~(\ref{eq.Approx_Javg}).
  Diamonds indicate the \emph{average} thresholds
  $\overline{k_c(a)}$ (solid symbols) and $\overline{a_c(k)}$ (open
  symbols) for arrangements of type 3.
  (b) Invasion threshold $k_c$ as
  a function of the average overlap $\langle {\cal{J}} \rangle$ for
   arrangements of
  type 1 (circles),  type 2  with orientation $\phi=0$ (diamonds)
  and type 3 with  mean orientation $\bar{\phi}=0$
  and two widths of uniform distribution, $\Delta \phi = 1$ (solid
  squares) and  $\Delta \phi =  2\pi$ (open squares).
  The branching host used as a motif for arrangements of type 2 and 3 is
  displayed  in the figure.
  The dashed line represents the dependence of $k_c^{0}$ vs $\langle
  {\cal{J}} \rangle$ given by Eq.~(\ref{eq.k_c0}).
  The insets show the dispersions $V_1$ and
  $V_2$ of the overlaps associated with the disorder and
  anisotropy, respectively, corresponding to the same arrangements as
  in the main figure (the
   symbol code is the same as in the main figure).}
\label{fig:Phase_1}
  \end{center}
\end{figure}

The invasion probability in host arrangements of type 1 is presented
in Fig.~\ref{fig:Dis_1}.  In particular, the inset shows the
dependence of the probability of invasion on the infection efficiency.
The value of $P_{\text{inv}}(k)$ is zero at small values of $k$ and
becomes positive above the threshold, $k> k_c(a)$.  As expected,
$k_c(a)$ increases with the distance between hosts (cf. the curves
marked by different symbols in the inset), $a$, since the infection
mechanism between neighbouring hosts should be more efficient in order
to invade the system with larger lattice spacing.  However, the curves
for $P_{\text{inv}}$ at different $a$ collapse onto a single master
curve (see Fig.~\ref{fig:Dis_1}), if plotted as a function of an
alternative control parameter, the average transmissibility, $\langle
{\cal{T}} \rangle$, which plays the same role as bond probability in
the percolation problem \citep{sander2002,sander2003}.  This statement
was first demonstrated heuristically by \cite{sander2002} for SIR
processes with heterogeneous transmissibility. Indeed, it is easy to
go beyond heuristic arguments and prove that the statement is valid if
the transmissibilities $\{T_{ij}\}$ are \emph{independent and
  self-averaging} quantities (see demonstration in
Appendix~\ref{App.MF_Hetero} and
\cite{kuulasmaa1982,Miller2007,Kenah2007} for cases in which the
transmissibilities are not independent and such description is not
appropriate). The collapse in Fig.~\ref{fig:Dis_1} suggests that
transmissibilities in arrangements of type 1 satisfy these conditions.
In fact, the master curve defined by the collapse coincides with that
obtained for the effective \emph{homogeneous} `mean-field' system with
transmissibility $\langle {\cal{T}} \rangle$ between all the nearest
neighbours (see the dashed curve in Fig.~\ref{fig:Dis_1} which is
identical for all lattice spacings).  This means that the
heterogeneous system is equivalent to a homogeneous `mean-field' one
for which the invasion threshold can be determined in an efficient
manner by solving the following equation,
\begin{equation}
\label{eq.Crit_Arr1}
\langle {\cal{T}} \rangle=p_c~,
\end{equation}
where $p_c\simeq 0.347$ is the bond percolation threshold in an
infinite (i.e., $L \rightarrow \infty$) triangular lattice
\citep{Isichenko_RMP1992,Stauffer1994}.  The value of $p_c$ provides
the minimally safe bound for invasion threshold in other 2D lattices
\citep{Isichenko_RMP1992,Stauffer1994}.  Since the average $\langle
{\cal{T}} \rangle$ converges very fast with $L$ to its limiting value
for $L \rightarrow \infty$, the value of $k_c(a)$ estimated from
Eq.~(\ref{eq.Crit_Arr1}) is representative for macroscopic systems.
The dependence of infection efficiency on lattice spacing defines the
phase boundary in the $(k,a)$ plane (see solid circles in
Fig.~\ref{fig:Phase_1}(a)) between the invasive and non-invasive
regimes. The threshold $a_c(k)$ provides the same separatrix.

%

%

The influence of the host
morphology on the invasion threshold can be better understood by
analysing the dependence of
$k_c$ on the overlaps ${\cal{J}}$.
The first term in Eq.~(\ref{eq.General_kc}) can be found by solving
Eq.~(\ref{eq.Crit_Arr1}) for the homogeneous system
in which all the overlaps are replaced by its mean value, $\langle
{\cal{J}} \rangle$, i.e.
\begin{equation}
\label{eq.k_c0}
k_c^0=\frac{|\ln(1-p_c)|}{\langle {\cal{J}} \rangle}~.
\end{equation}
It can be rigorously shown that the value of $k_c^0$
underestimates the threshold, i.e.  $k_c \geq k_c^0$ (cf. circles and
dashed line in Fig.~\ref{fig:Phase_1}(b)).  To prove this
  inequality we write the expression (\ref{eq.Crit_Arr1}) in terms of
  the overlap as $1-\langle e^{-k_c \J} \rangle =p_c$. Applying then
  the general inequality $\langle e^{-k_c \J} \rangle \geq e^{-k_c
    \langle \J \rangle}$ to the above relation we obtain $k_c \geq
  -\ln(1-p_c)/\langle \J \rangle$ which reduces to $k_c \geq k_c^0$
  after using the expression (\ref{eq.k_c0}) for $k_c^0$.
This inequality implies that the
contribution $\Delta k$ to $k_c$ associated with the dispersion of the
overlaps makes systems more resilient to epidemic invasion. An approximate solution
of Eq.~(\ref{eq.Crit_Arr1}) obtained by keeping the first correction
to $k_c^0$ gives the
low-bound estimate $\Delta k_1$  of $\Delta k$, i.e.
\begin{equation}
\label{eq.Dk_Arr1}
\Delta k \gtrsim \Delta k_1 = \frac{(k_c^0)^2}{2 \langle {\cal{J}} \rangle}V_1,
\end{equation}
which is proportional to the variance of the overlaps,
$V_1= \langle {\cal{J}}^2 \rangle-\langle {\cal{J}} \rangle^2$,
with other central moments of higher order being dropped (see the
derivation of Eq.~(\ref{eq.Dk_Arr1}) in Appendix \ref{App.Analytic_InvThres}).
The upper inset in
Fig.~\ref{fig:Phase_1} shows that the value of $V_1$ is non-zero,
as expected from the comparison of $k_c$ and
$k_c^0$ plotted in the main figure.

The low-bound estimate of $\Delta k$ given by Eq.~(\ref{eq.Dk_Arr1})
provides a safe threshold, $k_c^0+\Delta k_1$,
for the actual system. We expect this to be the case for a wide class
of morphologically complex hosts so that describing ${\cal{J}}$ in terms
of its average and standard deviation provides a minimally safe bound
to the invasion threshold in heterogeneous arrangements.

\subsection{Arrangements 2. Identical $n_i$ and $\phi_i$ at all the nodes}

\begin{figure}
\begin{center}


\begin{tabular}{l}
\large{(a)}\\
{\includegraphics[clip=true,width=9.0cm]{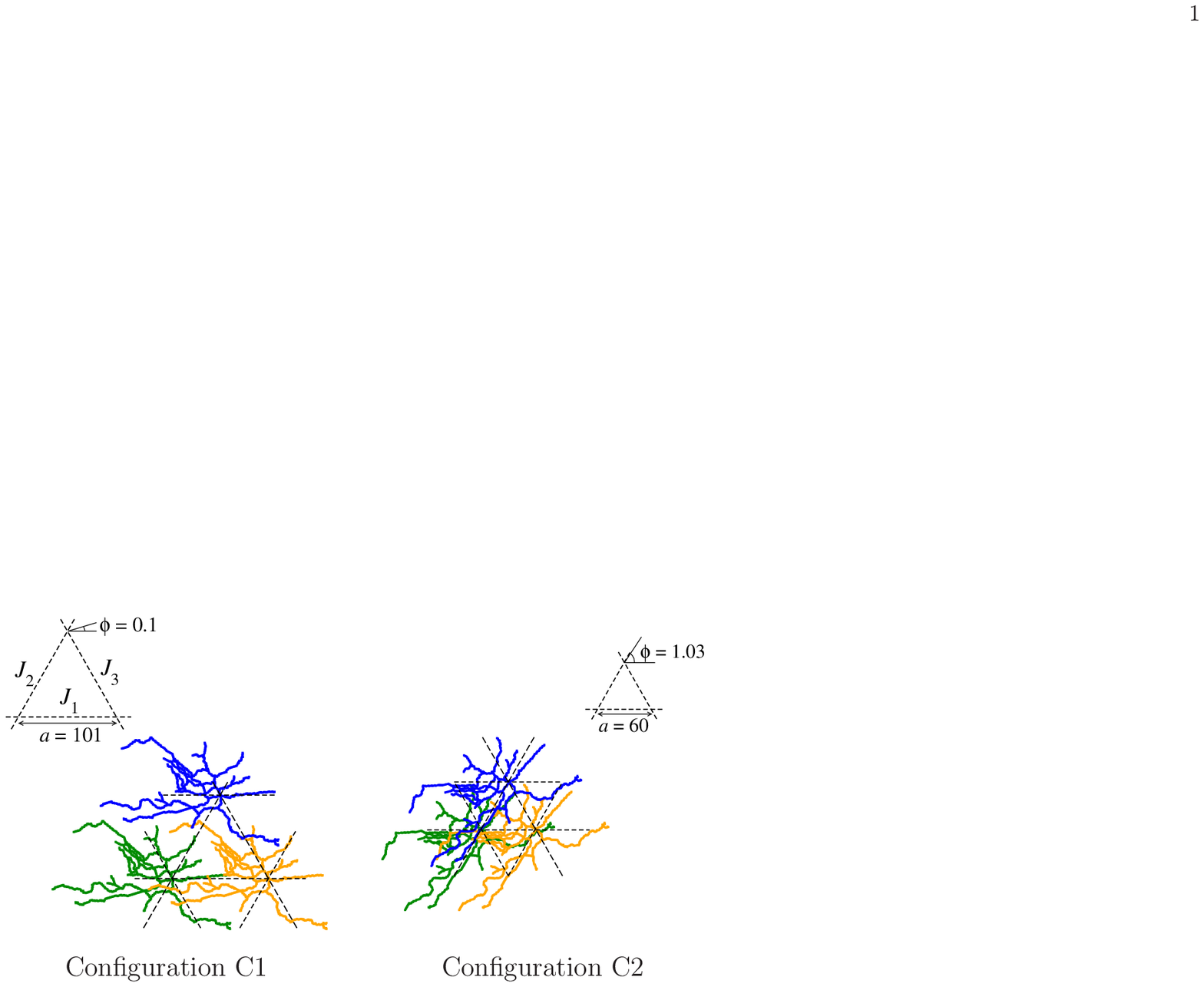}}
\end{tabular}
\begin{tabular}{l}
\large{(b)}\\
{\includegraphics[clip=true,width=9.0cm]{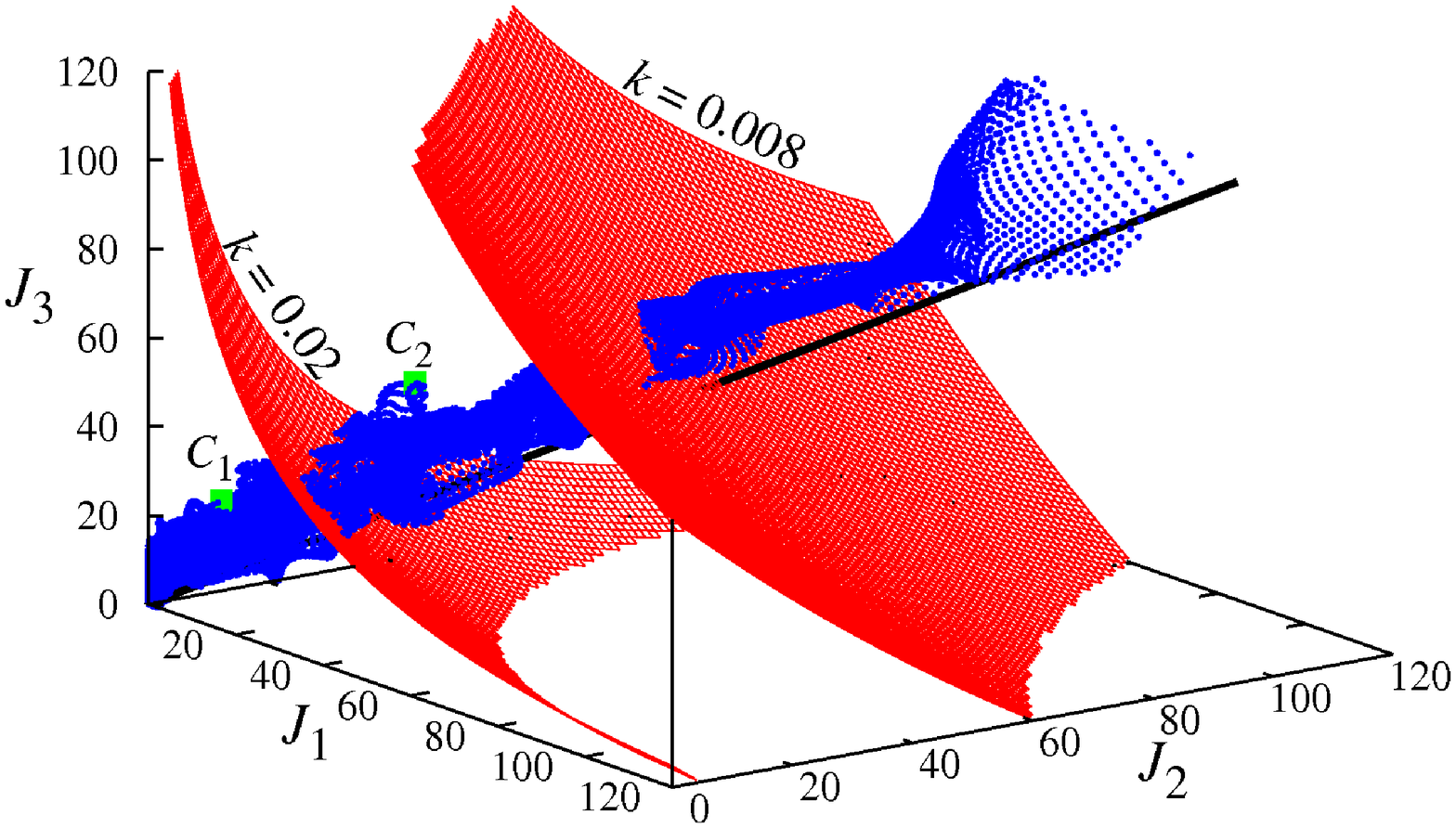}}\\
\end{tabular}


\caption{Arrangements of type 2 (an identical host with the same
  orientation placed at all the nodes).  (a) Unit cell of two
  different configurations constructed by using the same branching
  host.  The orientation, $\phi$, and spacing, $a$, corresponding to
  each case are indicated in the schematic reference frames displayed
  at the top. As shown in the left frame, there are three different
  values of the overlaps, $\{J_1,J_2,J_3\}$, corresponding to each of
  the main directions in the lattice.  (b) Space of overlaps,
  $(J_1,J_2,J_3)$. Each configuration with different $a$ and $\phi$ is
  mapped into a point in this space. The set of overlaps corresponding
  to all the possible configurations obtained for a given host defines
  its overlap locus (surface in blue).  The deviation of the blue
  surface from the straight black line representing overlaps between
  isotropic hosts ($J_1=J_2=J_3$) shows the degree of anisotropy in
  the overlaps for a typical branching host (shown in (a)).
  High values of the
  overlaps correspond to small lattice spacings $a$. In particular,
  the points labeled as C1 and C2 correspond to the configurations
  shown in (a).  The critical surface defined by $g(J_1,J_2,J_3,k)=0$
  (see Eq.~(\ref{eq:gEq0})) is shown in red for two values of the
  infection efficiency: $k=0.02$ and $k=0.008$. For a given value of
  $k$, the points in the overlap locus below/above the critical
  surface correspond to safe/vulnerable configurations. The
  intersection of the overlap locus with the critical surfaces
  parameterized by $k$ determines the critical threshold
  $a_c(k,\phi)$. The critical infection efficiency $k_c(a,\phi)$
  corresponding to a configuration with lattice spacing $a$ and
  orientation $\phi$ is given by the value of $k$ related to the
  critical surface containing the configuration point
  $\{J_1(a,\phi),J_2(a,\phi),J_3(a,\phi)\}$.}
\label{fig:J_Space}
\end{center}
\end{figure}

In ordered arrangements of type 2,  all the hosts are identical
branching structures with the same orientation.
The host overlaps along the main lattice directions,
$\{J_1,J_2,J_3\}$ (see Figs.~\ref{fig:J_Space}(a)),  depend on the lattice
spacing $a$, the host used as a motif, $n$, and its orientation,
$\phi$.
For a given $n$, the set of all possible overlaps obtained by
varying $a$ and $\phi$ within their respective domains define the host
\emph{overlap locus}. Fig.~\ref{fig:J_Space}(b) shows the
overlap locus corresponding to a typical branching host displayed
in Fig.~\ref{fig:J_Space}(a).
Each configuration with given $a$ and $\phi$ is represented by a point
$(J_1(a,\phi),J_2(a,\phi),J_3(a,\phi))$ belonging to the overlap
locus.
The inherent anisotropy of the host is reflected
in the dispersion of the overlaps, $J_1 \neq J_2 \neq J_3
\neq J_1$, and gives rise to a significant deviation of the overlap
locus (surface in blue) from the straight line (in black)
corresponding to the locus for isotropic host ($J_1=J_2=J_3$).

Similarly to arrangements of type 1, the spread of disease in ordered
arrangements can be mapped onto the bond percolation problem but now
with anisotropic bond probabilities corresponding to the values of
transmissibilities, $T_1\ne T_2 \ne T_3\ne T_1$, along the main
lattice directions defined as $T_i=1-e^{-k J_i}$ for $i=1,2,3$.
From this mapping, the invasion
threshold is defined by the following condition  \citep{Sykes_Essam1964},
\begin{equation}
\label{eq:anisotropic_p}
1 - T_1 - T_2 - T_3 + T_1T_2T_3=0~.
\end{equation}
This condition can be recast
in terms of the infection efficiency and overlaps,
as $g(J_1,J_2,J_3,k_c)=0$, where
\begin{equation}
\label{eq:gEq0}
\begin{split}
  g(J_1,J_2,J_3,k)&=1-e^{-k(J_1+J_2)}-e^{-k(J_1+J_3)}-e^{-k(J_2+J_3)}\\
  &+e^{-k(J_1+J_2+J_3)}.
\end{split}
\end{equation}
For given value of $k$, the condition $g=0$ defines a critical surface
(in red in Fig.~\ref{fig:J_Space}(b))  in the overlap space.
Therefore, the intersection of the critical surface with the overlap
locus (in blue) defines the invasion threshold, $a_c(k,\phi)$.
On the other hand, for a given value of $a$ and $\phi$,
the critical infection efficiency $k_c(a,\phi)$  is given by the
value of $k$ which generates a critical surface containing the point
$(J_1(a,\phi),J_2(a,\phi),J_3(a,\phi))$.

In practice, it is useful to consider the minimally resilient
thresholds $k^{\text{min}}_c(a)=\min_{\phi}\{k_c(a,\phi)\}$ or
$a^{\text{max}}_c(k)=\max_{\phi}\{a_c(k,\phi)\}$ which ensure that the
system is safe for any orientation if $k<k^{\text{min}}_c(a)$ or
$a>a^{\text{max}}_c(k)$, respectively. The invasion thresholds are
host-dependent and define different separatrices in the $(a,k)$ plane
for each host.  Fig.~\ref{fig:Phase_1}(a) shows the average thresholds
$\overline{k^{\text{min}}_c(a)}$ and $\overline{a^{\text{max}}_c(k)}$
over all the branching hosts $\{n \}$.  The phase boundary,
$\overline{a^{\text{max}}_c(k)}$, gives the safest estimate for the
invasion threshold, which is a consequence of the multivalued nature
of $a_c(k,\phi)$ in contrast to single-valued function $k_c(a,\phi)$
(see Appendix \ref{App.Multival_ac} for more detail).
%
%

The effects of the host anisotropy on the
invasion threshold can be analysed in a similar way as for
configurations of  type 1 by investigating
the dependence of $k_c$ on the overlaps.
Similarly, the critical infection efficiency, is given by
Eq.~(\ref{eq.General_kc}) with the mean-field value $k_c^0$ evaluated
for the system with mean overlap $\langle {\cal{J}} \rangle =
(J_1+J_2+J_3)/3$.
The dispersion in the overlaps results in the following approximate
expression for $\Delta k$ (see detailed derivation in Appendix \ref{App.Analytic_InvThres}),
\begin{equation}
 \label{eq.Dk_Arr2}
 \Delta k \gtrsim \Delta k_2 =  \frac{(k_c^0)^2}{2 \langle {\cal{J}} \rangle}V_2,
 \end{equation}
where the
quantity $V_2=\frac{1}{3(1+p_c)} \sum_{\alpha=1}^3 \left( J_{\alpha} -
  \langle {\cal{J}} \rangle \right)^2$ accounts for the anisotropy
of the overlaps.
The value of $\Delta k_2$ is non-negative meaning that $k_c \geq
k_c^0$, i.e. the anisotropy in host shape makes the system more resilient
as compared with the system of isotropic hosts with the same mean
overlap.

As an example, Fig.~\ref{fig:Phase_1}(b) shows that
$k_c>k_c^0$ in the
system of branching hosts, in agreement with the predicted
behaviour (cf. diamonds and the dashed line).
The inset shows the corresponding dispersion $V_2$. Given
that the deviation of $k_c$ from $k_c^0$ is small in the
system of branching hosts considered,
the correction $\Delta k_2$ is in fact a good approximation to
the actual deviation $\Delta k$.
This is the expected behaviour for systems of
hosts with moderate anisotropy.

\subsection{Arrangements 3. Identical $n_i$ and random $\phi_i$}

In the two previous sections, it has been shown that both the disorder
and anisotropy of the hosts make systems
of branching hosts more resilient
against epidemics. The arrangements of type 1 and 2 are extreme cases in the
sense that the first type highlights the morphological complexity ($V_1 \geq 0$
and $V_2=0$) whereas the second type highlights the effects of the
anisotropy of the hosts ($V_1=0$ and $V_2 \geq 0$). In generic
arrangements, such as those of the type 3 defined above, the two
effects are present. By assuming that the transmissibilities between
different pairs of hosts are independent, it is possible to show that
the behaviour of the actual heterogeneous
 system is equivalent to that of a
mean-field \emph{homogeneous} system with anisotropic
transmissibilities
$\{\langle {\cal{T}}_1 \rangle, \langle {\cal{T}}_2 \rangle, \langle
{\cal{T}}_3 \rangle \}$, where $\langle {\cal{T}}_{\alpha} \rangle$ is
the
average of the transmissibility along the direction $\alpha$ in the
lattice.
A proof for this statement, which represents a generalisation to
  systems with anisotropic transmissibilities of the mean-field
  description suggested in \cite{sander2002,sander2003}, is
  given in Appendix \ref{App.MF_Hetero}.
The equation for the invasion threshold in this case is
\begin{equation}
\label{eq:anisotropic_Avg_p}
1 - \langle {\cal{T}}_1\rangle - \langle {\cal{T}}_2 \rangle - \langle
{\cal{T}}_3 \rangle +
\langle {\cal{T}}_1 \rangle \langle {\cal{T}}_2 \rangle \langle
{\cal{T}}_3 \rangle=0,
\end{equation}
which generalises the formulae (\ref{eq.Crit_Arr1}) and
(\ref{eq:anisotropic_p}) in such a way that Eq.~(\ref{eq.Crit_Arr1})
corresponds to the particular case of Eq.~(\ref{eq:anisotropic_Avg_p})
 when $\langle {\cal{T}}\rangle$ is
the same along all the directions and Eq.~(\ref{eq:anisotropic_p})
emerges when there is no disorder in the anisotropic
transmissibilities.

As for the previous arrangements, the critical infection efficiency
obeys Eq.~(\ref{eq.General_kc}) with both heterogeneity and anisotropy
contributing to $\Delta k$.  The lowest-order approximation to the
lower bound for $\Delta k$ is given by the relation $\Delta k \gtrsim
\Delta k_3 = \Delta k_1 + \Delta k_2$ (see details in Appendix
\ref{App.Analytic_InvThres}) where $\Delta k_1$ and $\Delta k_2$ are
defined in Eqs.~(\ref{eq.Dk_Arr1}) and (\ref{eq.Dk_Arr2}),
respectively, with the dispersions terms generalised to
\begin{align}
V_1=&\frac{1}{3}\sum_{\alpha=1}^3 \left(\langle {\cal{J}}_{\alpha}^2\rangle - \langle
{\cal{J}}_{\alpha}\rangle^2\right) ,\\
V_2=&\frac{1}{3(1+p_c)}\sum_{\alpha=1}^3 (\langle {\cal{J}}_{\alpha}\rangle -
\langle {\cal{J}} \rangle)^2.
\end{align}
Fig.~\ref{fig:Phase_1}(b) shows
that the increase in orientational variability $\Delta \phi$ brings
additional heterogeneity in the system and thus results in a decrease
of $V_2$ but this does not necessarily
induce a decrease in
the value of critical infection efficiency
since the contribution $V_1$ may increase.
This illustrates the interplay between the role of disorder
and host anisotropy.

\subsection{Description of the invasion threshold in terms of
  morphological characteristics}

In the previous sections, we have established a link between the
invasion threshold and  host overlaps ${\cal{J}}$
characterised by the first moment, $\langle {\cal{J}} \rangle$, and the
deviations from the mean, $V_1$ and $V_2$.
The missing link between the host morphology and invasion threshold
can be recovered by studying how the morphology affects  overlaps.
Here we
show that both $\langle {\cal{J}} \rangle$ and the anisotropy of ${\cal
  J}$ (i.e. $V_2$) can be well described in terms of a reduced number of
morphological characteristics.
In contrast, a proper description of
the disorder-induced
contribution from $ \Delta k_1$
in terms of a reasonably small set of morphological characteristics is
hardly possible
and requires instead knowledge of the spatial host  density,
$\rho({\bf r})$ (see Eq.~(\ref{eq:density})).
However, we can ignore the disorder-induced contributions  from $ \Delta
k_1$ in order to obtain  a safe lower bound for critical infection efficiency
and thus connect this quantity with morphological characteristics of hosts.

We start the analysis by considering arrangements of type 1. The
overlaps ${\cal{J}}$ are statistically isotropic (i.e., $V_2=0$,
Fig.~\ref{fig:Phase_1}(b)) so that the invasion threshold can be
described in terms of a mean-field system with \emph{homogeneous} and
\emph{isotropic} overlaps, ${\cal J}^{\text{mf}}$,
and corresponding transmissibilities $\langle {\cal{T}} \rangle$.
The mean-field system consists of effective circular
hosts of radius $r^{\text{mf}}$ with overlap ${\cal J}^{\text{mf}}$.
These effective circles are fully described by the radius,
$r^{\text{mf}}$, and density, $\rho^{\text{mf}}(r)$, which is positive for
$0<r\le r^{\text{mf}}$ and is zero otherwise.
The functional form of $\rho^{\text{mf}}(r)$ is the same as of real branching
hosts in $[0,r^{\text{mf}}]$, i.e.  $\rho(r)\propto
r^{d_f-2}$ where $d_f$ stands for
the fractal dimension (see Appendix \ref{App.MF_Shapes} for more detail).
The radius of the effective circles is defined as
$r^{\text{mf}}=\frac{d_f+1}{d_f} \langle r \rangle$ to ensure
 that the average radius of the effective circles
coincides with the average radius of the actual hosts, $\langle r
\rangle$ (see Appendix \ref{App.MF_Shapes}).
Under these assumptions, the expression for the overlap between
neighbouring effective circles is given by
\begin{equation}
\label{eq.Approx_Javg}
\langle {\cal{J}} \rangle
\simeq
{\cal J}^{\text{mf}}
\simeq
\frac{1}{3}\left(\frac{M
    d_f}{\pi a}\right)^2
 \left(\frac{r^{\text{mf}}}{a}\right)^{-7/2}
\left(\frac{2r^{\text{mf}}}{a}-1 \right)^{3/2}~,
\end{equation}
valid for small overlaps when $2r^{\text{mf}}/a-1 \ll 1$.  The
parameter $M$ is a normalisation constant (see Appendix
\ref{App.MF_Shapes}).  The substitution of this expression for $
\langle {\cal{J}} \rangle $ into Eq.~(\ref{eq.k_c0}) gives the
required link between the invasion threshold and the effective radius,
$k_c^{\text{mf}} \propto
(r^{\text{mf}}/a)^{7/2}(2r^{\text{mf}}/a-1)^{-3/2}$.  In fact, for the
particular system of branching hosts studied here, this is a very good
estimate for the value of the critical infection efficiency, $k_c
\gtrsim k_c^{\text{mf}}$, as demonstrated in Fig.~\ref{fig:Phase_1}
(the solid circled line falls inside the shaded gray area representing
the set of possible values for $k_c^{\text{mf}}$ within statistical
errors).  This means that for the arrangements of type 1, the
mean-field homogeneous system of effective circles can be reliably
used for estimating the phase boundaries.

\begin{figure}
\begin{center}


{\includegraphics[clip=true,width=8.4cm]{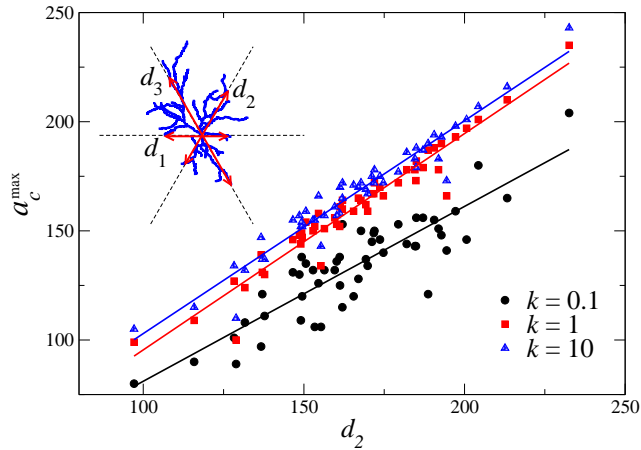}}


\caption{
Dependence of the critical lattice spacing,
$a_c^{\text{max}}(k)$ on the SLAD, $d_2$, for all the branching hosts
in arrangements of type 2.
Different symbols refer to different values of  $k$ as marked and
each point for a particular symbol corresponds to an individual
branching host.
The solid lines represent the linear regression fit for each value of
$k$ (e.g. $a_c(k)= 14.1 + 0.94 d_2 $ with correlation
coefficient $\simeq 0.97$ for $k= 1$).
 The inset defines graphically the lattice-adapted
  diameters $\{d_1,d_2,d_3\}$.
}
\label{fig:Phase_SLAD}
\end{center}
\end{figure}

In the case of ordered (type 2) and partially ordered (type 3)
arrangements, the overlaps along distinct lattice directions can be
significantly different and thus crucial for evaluation of the
invasion threshold in contrast to the mean overlap which is important
for disordered (mean-field like) arrangements of type 1.  Therefore,
instead of a single characteristic such as mean radius of the
effective circles for disordered arrangements, we introduce an ordered
set of linear sizes of branching hosts along the lattice directions,
i.e the set of lattice-adapted diameters, $\{d_1,d_2,d_3\}$ ($d_1 \le
d_2 \le d_3$; see illustration in Fig.~\ref{fig:Phase_SLAD}).  The
second lattice-adapted diameter (SLAD), $d_2$, plays the most important
role for finding the critical value of, e.g. lattice spacing, and thus
estimating the invasion threshold.  This is due to the fact that
Eqs.~(\ref{eq:anisotropic_p}) and (\ref{eq:anisotropic_Avg_p}) have a
solution for the critical threshold only if the overlaps at least
along two directions are finite.  The overlaps become greater
  than zero if the
lattice-adapted diameters are comparable or greater than the lattice
spacing and thus solution of Eqs.~(\ref{eq:anisotropic_p}) and
(\ref{eq:anisotropic_Avg_p}) exists if $a \lesssim d_2 $.  This
qualitative analysis suggests the existence of strong correlations
between the SLAD and $a_c$.  Indeed, we have observed such
correlations, i.e.  $a_c \simeq d_2$, for ordered arrangements of
type 2 (see Fig.~\ref{fig:Phase_SLAD}).  The description of $a_c$ in
terms of SLAD gets worse for small values of $k$ (see the black
squares in Fig.~\ref{fig:Phase_SLAD}) when the strong overlaps between
hosts should be achieved at criticality and thus the interior density
of the hosts, rather than the diameter only, becomes important.

\section{Discussion and Conclusions}
Using the framework of an SIR epidemiological model, we have
investigated the transmission of infection and spread of disease in systems of hosts with
realistically complex morphology.  Our main finding is that the greater the
irregularity
in the host morphology, the more resilient is the
population to epidemic invasion under otherwise identical conditions
(for instance, identical spatial arrangement of hosts).   We derive a safe lower bound for the invasion threshold, which has been obtained
for branching hosts with independent transmission rates placed on a
triangular lattice.
We have shown mathematically that
this bound holds for all other 2-D topological arrangements of hosts
with nearest-neighbour transmission and
even in the case when the transmission rates between different hosts
are correlated with each other.
In particular, irregularity in the host positions,
  i.e. small random displacements of hosts from lattice nodes bring
  correlations in transmissibilities and thus make the system more
  resilient. Of course, for some real systems, the assumption about
  nearest-neighbor transmission may be violated by the presence of
short-cuts between remote nodes due, for example, to wind or animal motion. In this
case, it is known  that the system becomes less
resilient to epidemic invasion \citep{sander2002} and the bounds given above are no longer
valid.

We have used a set of planar neurons to illustrate the effects of
  complex branching hosts on the spread of infection. We have
    identified two sources of heterogeneity in the systems considered:
    (i) morphological complexity of hosts and (ii) the host shape
    anisotropy.  Both contribute to the resilience of the system
    against epidemic invasion and can be described 
    by means of two
morphological
characteristics of hosts, i.e. by the mean effective
radius and the second lattice-adapted diameter. Such characterisation is
not exact in general but it provides a safe bound to the invasion threshold. The main conclusions are generic and remain valid for any type of morphologically complex hosts. In particular, the methodology introduced here and the bounds for resilience to invasion apply to the transmission of infection in other 2-D systems with analogous disorder expressed in the host morphology and anisotropy. Examples include the spread of plant disease through contacts between adjacent plants in a field, orchard or forest in which the host plants frequently  occur on a 2-D lattice. Here the contact structures between nearest-neighbours are determined  by overlap of shoots (for aerial pathogens) or roots (for soil-borne pathogens) \citep{Gilligan2008} in which the 3-D structure of the plant can be collapsed onto a 2-D framework when considering transmission of infection between nearest-neighbours.  In principle, a similar analysis could be performed for ensembles of morphologically complex 3-D hosts arranged on a 2-D lattice that takes explicit account of the three dimensional host structure. Expressions for the invasion threshold are not known analytically in this case and the precise definition of  quantities directly linked to host morphologies such as fractal dimensions or average radius  are system-dependent that require further study. Other potential applications of the methods include analysis of the transmission of infection via the `morphology' of contacts between clusters of susceptible hosts in social networks, in which the clusters can be approximated by a 2-D lattice (cf recent work on percolation models for the spread of plague through gerbil populations in lattices of interconnecting burrows \citep{Davis_Nature2008}).

The present work opens several possible directions of  further
  research for practical applications in considering disease control strategies and for basic understanding of epidemic spread involving heterogeneous transmission of infection.
For instance,
our analysis suggests new ways for control of epidemics
in real systems where host morphology is inherently complex.  For
example, in a system where an epidemic is active, a treatment
enhancing the anisotropy in the transmission rates, would be more
efficient as compared with reduction of transmission rates in all
directions, i.e. isotropically. Such might arise with the deployment of microbiological biological control agents to restrict the spread of infection of soil-borne pathogens in plant populations \citep{Gibson1999}. Biological control agents often exhibit marked variability in performance \citep{Gibson1999}, failing to provide isotropic control of infection on targeted hosts, for example due to uneven colonization of roots by microbial antagonists deployed as biological control agents. Our results suggest that increasing the degree of anisotropy due to these organisms may yet contribute to success in controlling invasion. Anisotropy may also be fostered in social or animal systems by preferential treatment of some, rather than all, connections between clusters.
The findings in the current manuscript are also important for
better understanding virus tracing of neurons (e.g.~\cite{Loewy:1998}), which is
a biological staining method where virus
propagation from neuron to neuron is used as a means to
histologicaly mark the interconnections, so that they become visible to the
microscope.   More specifically, our results imply that
virus tracing might be not so effective in the case of anysotropic
or not so complex neuronal cells, which could therefore be
overlooked by this type of marking.  Analogue effects can be also important
in transneuronal spreading of virus in order to deliver gene
therapy~\citep{Oztas:2003}.

Finally, the approach presented here is relevant to epidemics for
which an SIR model is suitable and a mapping to ordinary percolation
exists. While many diseases can be described by the SIR framework,
others cannot. It would be interesting to analyze the effect of host
morphology within the framework of a different family of
epidemiological models with final state not immune to the disease
(such as the susceptible–infected–susceptible model). Such an
extension is challenging since the mapping to ordinary percolation is
no longer possible requiring tools such as directed percolation
\citep{Marro_99:book,Hinrichsen_00} from non-equilibrium physics.

\section*{Appendices}
\appendix

\section{Homogeneous recovery times}
\label{App.Homo_tau}
In the present work, we have assumed that  the recovery times of the
hosts are homogeneous.
This assumption provides a minimally safe bound for
the invasion threshold for epidemic outbreak  so that the systems with
disorder in $\tau$  are safer i.e. less likely to be invaded.
Indeed, the probability of invasion
$P_{\text{inv}}^{\mathcal{A}}$ in an SIR process $\mathcal{A}$ with
heterogeneous $\tau$ and average transmissibility $\langle {\cal{T}}
\rangle$ satisfies \citep{kuulasmaa1982,Cox1988}:
\begin{equation}
\label{eq.Bounds.1}
P_{\text{inv}}^{\mathcal{A}} \leq P_{\text{inv}}^{\mathcal{B}},
\end{equation}
where $P_{\text{inv}}^{\mathcal{B}}$ is the probability of invasion in an SIR
process ${\cal B}$ with the same average transmissibility as in
$\mathcal{A}$ but with homogeneous $\tau$.
The above inequality
implies that the non-invasive region for the
heterogeneous (in $\tau$) system is wider than for the homogeneous one.

\section{Validity of the mean-field description of
  heterogeneous systems}
\label{App.MF_Hetero}
\begin{figure}[h]
\begin{center}
{\includegraphics[clip=true,width=11cm]{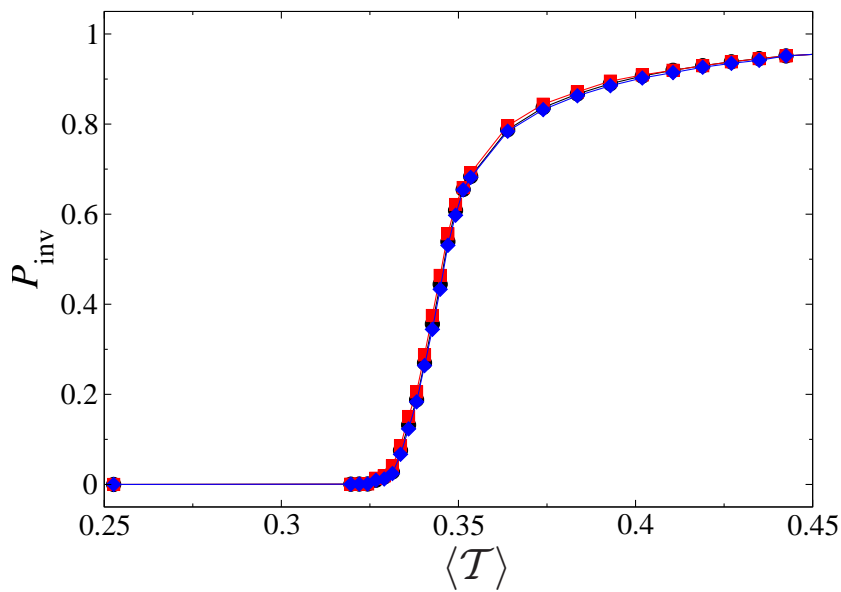}}
\caption{Probability of invasion, $P_{\text{inv}}$, for arrangement of
  type 1 (system 
  size $L\times L  = 200\times 200$) with lattice spacing $a=80$. 
Different symbols (squares, circles, and diamonds) 
correspond to three different  configurations of
hosts.}
\label{fig.Pinv_SeveralConfs}
\end{center} 
\end{figure}

The validity of the mean-field description leading to the threshold
condition (2) in the main text in systems with isotropic disorder was
heuristically suggested by \cite{sander2002}.  In fact, it can be
proven more generally that \emph{the mean-field description is valid
  both in the presence of isotropic and anisotropic uncorrelated
  disorder provided the overlaps between different pairs of hosts (and thus
  transmissibilities) are
  self-averaging independent quantities\footnote{A physical
    characteristic of a disordered system is said to be self-averaging
    if its average over several configurations of disorder coincides
    with its average in a single infinitely large configuration
    \citep{Sornette2000}.}.} In order to prove this, let us interpret
$T_{ij}$ as the \emph{conditional} probability,
$T_{ij}(J)=P(i\rightarrow j | J)$, for infection to be transmitted
from host $i$ to host $j$ \emph{given} the overlap $J$ which is a
random variable.  In the anisotropic case, the overlaps along
different lattice directions are distributed according to distinct and
independent probability densities, e.g., $\{f_\alpha(J),\alpha= 1,2,3
\}$ for a triangular lattice.  The probability that the disease is
transmitted from an infected host at node $i$ to one of its neighbors
at node $j$ along the direction $\alpha$ is given then by the
following expression,
\begin{equation}
\langle T^{(\alpha)}_{ij} \rangle = \int_J T_{ij}(J) f_{\alpha}(J) \text{d}J
\label{eq:mapping}
\end{equation}
which is identical for all the pairs $(i,j)$ along direction $\alpha$
and thus coincides with the average transmissibility $\langle {\cal
  T}_{\alpha} \rangle \equiv \langle T^{(\alpha)}_{ij} \rangle$ if the
transmissibility is a self-averaging quantity. We have
checked that the self-averaging condition indeed holds for branching
structures by demonstrating that the probability of invasion does not
depend on the particular configuration of hosts.  This is illustrated
in Fig.~\ref{fig.Pinv_SeveralConfs} where we show that the invasion
probability curves ($P_{\text{inv}}$ vs $\langle {\cal T} \rangle $)
collapse for different realizations of disorder.

As a consequence of the above analysis, the system can be mapped onto
an anisotropic but homogeneous mean-field system with
transmissibilities, $\{ \langle {\cal T}_{\alpha} \rangle \}$.
Therefore, Eq.~(8, main text) giving the invasion threshold for an
anisotropic disordered system can be obtained from Eq.~(5, main text)
by replacing the anisotropic transmissibilities $\{ T_{\alpha}\}$ by
the mean values $\{\langle {\cal T}_{\alpha} \rangle \}$.  In the
particular case of isotropic triangular system, $\langle {\cal
  T}_{\alpha} \rangle = \langle {\cal T} \rangle $, so that Eq.~(8,
main text) reduces to $1-3\langle {\cal T} \rangle+\langle {\cal T}
\rangle^3=0$ which has the solution $\langle {\cal T} \rangle=p_c$,
where $p_c=2 \sin(\pi/18) \simeq 0.347$
\citep{Isichenko_RMP1992,Stauffer1994}.

\section{Multivalued  behaviour of critical lattice spacing }
\label{App.Multival_ac}
\begin{figure}[h] 
\begin{center}
{\includegraphics[clip=true,width=12cm]{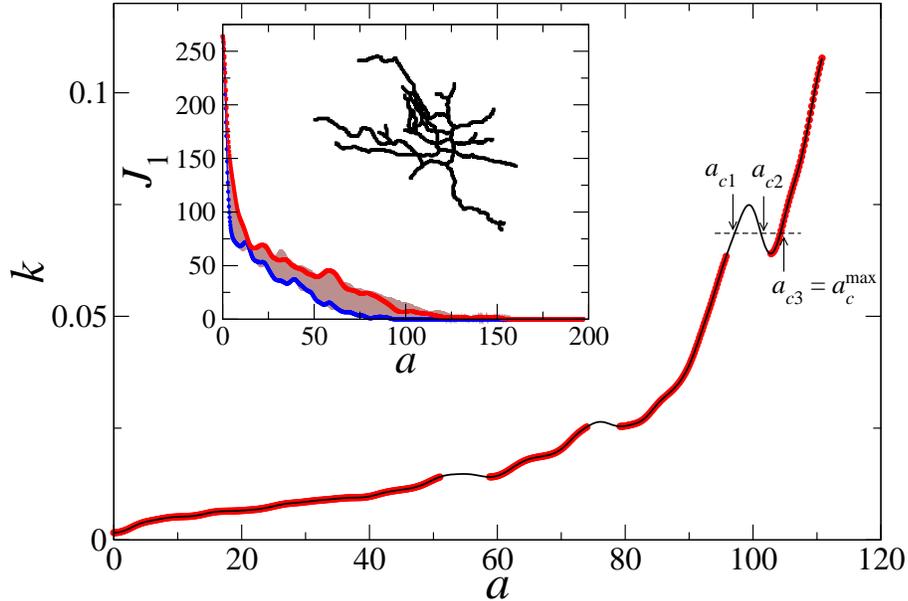}}
\caption{
  Infection efficiency versus lattice spacing for fixed
  orientation of branching hosts represented by a planar neuron shown
  in the inset. 
The definition of the maximally resilient threshold
$a^{\text{max}}_c(k)$ (red curve) is illustrated for a particular
value of $k$ for which $a_c(k,\phi)$ takes three different values.
%
%
  The inset shows the dependence $J_1(a)$ along one of the lattice
  directions for the arrangement of type 2 obtained by using the
  displayed host as a motif.  The shaded (brown) region indicates the
  overlaps for all the possible orientations, $\phi$, of the
  host. Blue and red lines correspond to $J_1(a)$ for $\phi=0$ and
  $\phi=1$, respectively. }
\label{fig.k_vs_a_overlap}
\end{center} 
\end{figure}

The invasion threshold in the ordered arrangements of type 2 (the same
host is placed on all the nodes with the same orientation, $\phi$) is
given by the condition $g=0$ (see Eq.~(6) in the main text).  The
function $g$ depends on the overlaps along the main lattice
directions, $\{J_\alpha(a,\phi),\alpha=1,2,3\}$ .  For fixed
orientation $\phi$, the overlap $J_\alpha(a,\phi)$ between branching
structures is generally a non-monotonic function of the lattice
spacing (see blue and red curves in the inset in
Fig.~\ref{fig.k_vs_a_overlap}).  In this figure, we show that the
non-monotonic character of the overlaps results in non-monotonic
behaviour of the critical infection efficiency, $k_c(a,\phi)$, and
confers a multivaluated dependence of the critical lattice spacing,
$a_c(k,\phi)$ versus $k$.  As a consequence, the minimally resilient
threshold in terms of the lattice spacing,
$a^{\text{max}}_c(k)=\max_{\phi}\{a_c(k,\phi)\}$ (see the illustration
for $a^{\text{max}}_c(k)$ in Fig.~\ref{fig.k_vs_a_overlap} where this
quantity is represented by the red curve) gives a safer boundary as
compared to the minimally resilient threshold in terms of the
infection efficiency, $k^{\text{min}}_c(a)=\min_{\phi}\{k_c(a,\phi)\}$
(the black curve), i.e. the red curve in Fig.~\ref{fig.k_vs_a_overlap}
either coincides with or is to the right of the black one.

\section{Analytical estimates for the invasion threshold}
\label{App.Analytic_InvThres}
In this Section, 
we derive an approximate analytical expression for $k_c$ for
arrangements of type 3 and extend this result to other arrangements.  



We start from
Eq.~(8) in the main text which is valid for 
%
arrangements of 
type 3. 
Expressing the transmissibilities, $\{\T_{\alpha},\alpha=1,2,3\}$, in
terms of anisotropic overlaps, $\{\J_{\alpha},\alpha=1,2,3\}$, gives 
\begin{equation}
\label{eq:Threshold_1}
1-\langle e^{-k_c\J_1} \rangle \langle e^{-k_c\J_2} \rangle
-\langle e^{-k_c\J_1} \rangle \langle e^{-k_c\J_3} \rangle
-\langle e^{-k_c\J_2} \rangle \langle e^{-k_c\J_3} \rangle
+\langle e^{-k_c\J_1} \rangle \langle e^{-k_c\J_2} \rangle 
\langle e^{-k_c\J_3} \rangle=0~.
\end{equation}
The overlaps, $\J_{\alpha}=\langle\J_{\alpha}\rangle +
\delta\J_{\alpha} $, along lattice direction $\alpha$ are random
values characterized by the mean value $\langle\J_{\alpha}\rangle$ and
deviations from the mean, $\delta\J_{\alpha}$.  Anisotropy brings an
additional source of dispersion, $ \Delta \J_{\alpha} = \langle
\J_{\alpha} \rangle - \langle \J \rangle $, where the overall mean
overlap is $\langle \J \rangle = \frac{1}{3} \sum_{\alpha=1}^3 \langle
\J_{\alpha} \rangle$.  Expanding Eq.~(\ref{eq:Threshold_1}) in small $
\delta \J_{\alpha} \ll \langle \J_\alpha \rangle$ first and then in
$\Delta \J_{\alpha} \ll \langle \J \rangle $ and keeping the leading
corrections to the the mean-field value, results in the following
expression:
\begin{equation}
\label{eq:Threshold_2}
1-3 e^{-2k_c \langle \J \rangle}+ e^{-3k_c \langle \J
  \rangle}+
\left[ \left(\frac{e^{-k_c \langle \J \rangle}}{2} -1 \right) \sum_{\alpha=1}^3 \langle (\delta \J_\alpha)^2
\rangle - \frac{1}{2}
  \sum_{\alpha=1}^3(\Delta \J_{\alpha})^2 \right] k_c^2 e^{-2k_c \langle
  \J \rangle}\simeq 0~.
\end{equation}

In order to derive an analytical estimate for the critical infection
efficiency from Eq.~(\ref{eq:Threshold_2}), it is convenient to
separate the mean-field contribution, 
$k_c^0$, 
i.e. 
$
k_c = k_c^0 +\Delta k 
$,  
and evaluate the value of $\Delta k$ by expanding  
Eq.~(\ref{eq:Threshold_2}) in $\Delta k \ll k_c^0$. 
The zero-order term gives the condition for the critical mean-field
value of the infection efficiency, 
\begin{equation}
1-3 e^{-2k_c^0 \langle \J \rangle}+ e^{-3k_c^0 \langle \J
  \rangle} = 0~, 
\end{equation}
which has the solution 
$k_c^0=-\frac{\ln (1-p_c)}{\langle \J  \rangle}$, 
where $p_c=2 \sin(\pi/18) \simeq 0.347$.

The next-order terms give an estimate for $\Delta k$, 
\begin{equation}
\Delta k \gtrsim \Delta k_3 = 
\frac{(k_c^0)^2}{2 \langle {\cal{J}} \rangle}(V_1+V_2)~,
\end{equation}
where $V_1$ and $V_2$ are defined in the main text by Eqs.~(9) and
(10).

The estimates of $\Delta k_1$ and  $\Delta k_2$ for arrangements of
type 1 and 2, respectively, can be obtained in a similar way 
 as particular cases of the derivation given above. 

\section{Mean-field description of complex shapes}
\label{App.MF_Shapes}
In this section, we show how the complex shapes can be approximately
represented  by effective regular shapes, i.e. effective circles. 

In disordered arrangements of type 1, an
arbitrary host $n_i$ from the set of nodes $\{n\} $ ($n=1,\ldots,N)$ is placed at each
lattice node $i$  at uniformly random orientation $\phi_i$. 
The invasion threshold can then be described
in terms of a mean-field system in which all the hosts have the
same density, $\langle \rho \rangle(r)$, where 
\begin{equation}
\label{eq.Dis_3}
  \langle \rho \rangle(r)= \frac{1}{N} \sum_{n=1}^N
  \int_{0}^{2\pi} \frac{\rho({\bf r};n)}{2\pi r} \text{d}\phi~, 
\end{equation}
defined as the average of the host density $\rho({\bf r};n)$ over the
$N$ morphologically different hosts in the ensemble and their
orientation. 
Such averaging eliminates the dependence on the
orientation so that $\langle \rho \rangle(r)$ depends on the radial
distance, $r=|{\bf r}|$, only. 
By definition, the integral of the density over the 2D plane gives the 
average ``mass'' of hosts, i.e. the mass of the effective host in the
mean-field system, 
%
\begin{equation}
M = \int_{\R^2} \langle \rho \rangle \text{d}^2r~.  
\end{equation}
The mean radius of the effective host is then given by 
\begin{equation}
\langle r \rangle=\int_{\R^2} r 
\langle \rho
\rangle \text{d}^2r~.
\label{eq:mean_r}
\end{equation}

\begin{figure}[h]
\begin{center}
{\includegraphics[clip=true,width=12cm]{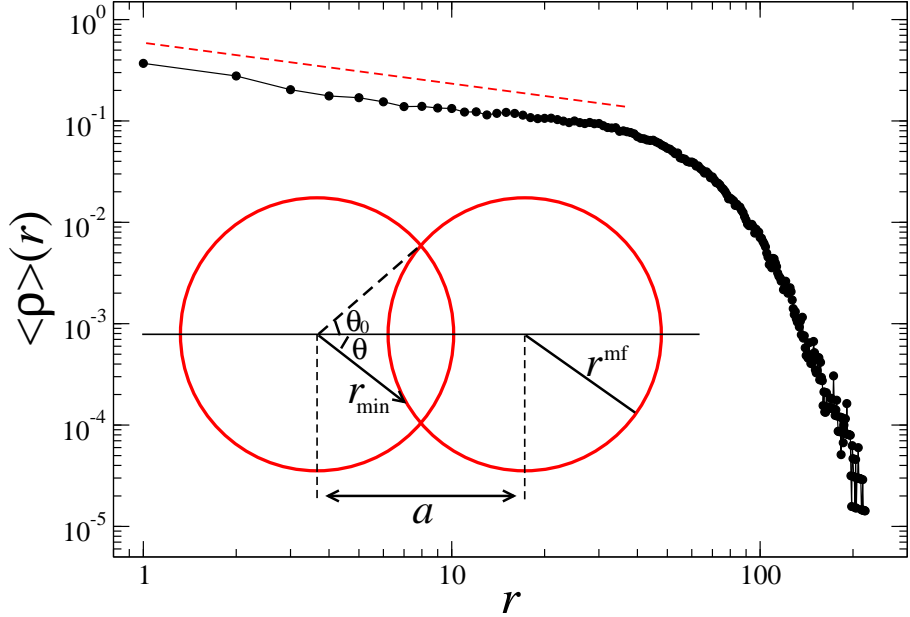}}\\
\caption{ Average density for branching structures for disordered
  arrangements of type 1 versus $r$.  The dashed line shows a pure
  power-law behaviour with exponent $d_f-2=-0.5$.  The inset
  illustrates the overlap of two effective circles of radius,
  $r^{\text{mf}}$, placed at a distance $a$ apart from each other. The
  overlap region is defined by the following conditions in polar
  coordinates $(r,\theta)$: $\{\theta \in [\theta_0,-\theta_0];r\in
  [r_{\text{min}}(\theta),r^{\text{mf}}]\}$.  }
\label{fig:rho_r_avg} 
  \end{center} 
\end{figure}

The  mean density can be easily obtained 
by numerical integration of Eq.~(\ref{eq.Dis_3}). 
Fig.~\ref{fig:rho_r_avg} shows that the mean density 
%
for branching structures
appears to
decay with radius according to a power law,  
$\langle \rho \rangle \sim r^{d_f-2}$ with $d_f=1.5 \pm
0.1$ (truncated at large values of $r$ by an exponential cut-off which
accounts for the finite radius of the neurons used here for illustration). 
This is the expected behaviour for branching structures  which are non-compact
objects and are then characterized by a fractal dimension $d_f$
smaller than dimensionality $d=2$ of the embedding space.

The exponential decay of the mean density at large values of $r$ can be
ignored in the first approximation so that the  hosts in the
mean-field description can be represented by  effective circles
of finite radius $r^{\text{mf}}$ with density, $\rho^{\text{mf}}(r)$,
given by 
\begin{equation}
\label{eq.Dis_5}
\rho^{\text{mf}}(r)=\begin{cases}
M \frac{d_f (r^{\text{mf}})^{-d_f}}{2 \pi} r^{d_f-2},& r \leq r^{\text{mf}}\\
0, & r > r^{\text{mf}}~.
\end{cases}
\end{equation}
The mean-field radius is defined by the following condition, 
\begin{equation}
 \langle r \rangle=\int\limits_0^{2\pi}\text{d}\theta \int\limits_0^{r^{\text{mf}}} 
r \rho^{\text{mf}}(r)~r\text{d}r = \frac{d_f}{d_f+1} r^{\text{mf}} 
~.   
\label{eq:r_mean}
\end{equation}
The value of the mean radius $\langle r \rangle$ in
Eq.~(\ref{eq:r_mean}) can be evaluated for the heterogeneous system
using 
Eq.~(\ref{eq:mean_r}) and thus used for definition of
$r^{\text{mf}}$ to ensure that the mean radius of the effective circle 
coincides with the mean radius of actual hosts. 

The overlap between two effective hosts represented by mean-field
circles  placed at a distance $a$ apart from
each other (see inset in Fig.~\ref{fig:rho_r_avg})
 is given by Eq.~(12) in the main text, i.e by the integral, 
\begin{equation}
\J^{\text{mf}} = \left(\frac{M
d_f}{2 \pi (r^{\text{mf}})^{d_f} }
\right)^2 \int\limits_{-\theta_0}^{\theta_0} \text{d} \theta 
\int\limits_{r_{\min}(\theta)}^{r^{\text{mf}}} r^{d_f-1}
\left(\sqrt{r^2+a^2-2ar \cos \theta} \right)^{d_f-2}\text{d}r~, 
\end{equation}
where $\theta_0$ and $r_{\min}$ are defined in
Fig.~\ref{fig:rho_r_avg}. 
If the overlap is small,  the above integral can be approximately
evaluated by performing an expansion in powers of the linear overlap 
$\delta = 2r^{\text{mf}}/a-1 \ll 1$, giving 
\begin{equation}
\J^{\text{mf}} = \left[\frac{1}{3}\left(\frac{M
    d_f}{\pi a}\right)^2\right] 
 \left(\frac{r^{\text{mf}}}{a}\right)^{-7/2}
\delta^{3/2} + {\cal O}(\delta^{5/2})~.
\end{equation}
In the system of
branching hosts used for illustration in the main text, $M =1282.4$
stands for the average number of pixels in the digital images,
$d_f=1.5 \pm 0.1$, and $\langle r \rangle = 50 \pm 5$. The effective
radius is then $r^{\text{mf}}=83 \pm 10$ and the approximation to the
overlap becomes $\J^{\text{mf}} \simeq (0.024 \pm 0.008)
\delta^{3/2}$.

\section*{Acknowledgments}
  FJPR, SNT, FMN and CAG thank BBSRC for funding (Grant
  No. RG46853). CAG also acknowledges support of a BBSRC Professorial
  Fellowship.  The authors are grateful to Prof. David Schubert (Salk
  Institute) for comments on disease spreading among neurons. LDFC
  thanks CNPq (308231/03-1) and FAPESP (05/00587-5) for sponsorship.
  Part of this work was performed during a Visiting Scholarship of
  LDFC to St. Catharine's College, University of Cambridge.

\bibliographystyle{JRSstyle}

\end{document}